\documentclass[aps,prd,twocolumn,superscriptaddress]{revtex4}
\usepackage{graphicx}
\usepackage{epsfig,epsf}
\usepackage{amsmath}
\usepackage{amsthm}
\usepackage{amsfonts}
\usepackage{amssymb}
\usepackage{dsfont}
\usepackage{multirow}
\usepackage{appendix}
\usepackage{slashed}
\usepackage[active]{srcltx}
\usepackage{psfrag}
\usepackage{subfigure}
\usepackage{capt-of}
\usepackage[utf8]{inputenc}
\begin{document}

\title{Impact of a thermal medium on newly observed $Z_{cs}(3985)$ resonance and its $ b $-partner}
\date{\today}
\author{J.Y.~S\"ung\"u}
\affiliation{Department of Physics, Kocaeli University, 41001
	Izmit, Turkey}
\author{A. T\"{u}rkan}
\affiliation{\"{O}zye\u{g}in University, Department of Natural and
	Mathematical Sciences, \c{C}ekmek\"{o}y,Istanbul,Turkey}
\author{H.~Sundu}
\affiliation{Department of Physics, Kocaeli University, 41001
	Izmit, Turkey}
\author{E. Veli Veliev}
\affiliation{Department of Physics, Kocaeli University, 41001
	Izmit, Turkey}
\begin{abstract}
Motivated by the very recent discovery of the strange hidden-charm exotic state $Z_{cs}(3985)$
by the BESIII Collaboration, we study the possible interpretation of this exotic state both at $ T= 0 $ and $ T\neq 0 $. We analytically compute the mass and meson-current coupling constant of this resonance with spin-parity $ J^{P} = 1^+$ at
finite temperature approximation up to the sixth order of the thermal operator dimension including non-perturbative contributions. Extracting thermal mass
and meson-current coupling constant sum rules, the modifications on
properties of $Z_{cs}(3985)$ state in hot medium is determined.  As a by-product, the hadronic parameters of the bottom partner of  $Z_{cs}(3985)$ is estimated as well. The search of temperature effects on the hadronic parameters of hidden-charm meson $Z_{cs}(3985)$ and the bottom partner make us understand the phase transitions, chiral symmetry breaking, and the properties of hot-dense matter in QCD. Moreover, the full
width of the resonance $Z_{cs}(3985)$ is calculated as $(12.0\pm 0.8)~\mathrm{MeV}$ using the strong decay in the tetraquark picture.  Results for width and mass are in reasonable agreement with existing experimental data, and results of other
theoretical works. The obtained information about the parameters of considered states is useful for experimental investigations of exotic mesons.
\end{abstract}
\maketitle
\section{Introduction}

Though many new exotic states named as XYZ states above $ D\overline{D} $ threshold have recently been observed in different experiments, their substructures can not be explained yet clearly. There are a lot of candidates for exotic hadrons in the charmonium sector of Quantum Chromodynamics (QCD) studied at vacuum, hot medium and also in nuclear medium, such as $ X (3872), Z_ c (3900), Z_ c (4430),Y(4260), Z_ b (10610),$ and $Z_{b}^{\prime} (10650) $ \cite{Brambilla:2019esw,Liu:2019zoy,VeliVeliev:2018eaw,Azizi:2020itk,Isik:2020fwl,Chen:2016qju,Azizi:2020yhs,Agaev:2016dev,Agaev:2017tzv,Agaev:2017lmc,Agaev:TJP,Ozdem:2017jqh,Hay:TJP} which yields a new horizon for understanding the inner structure of strongly interacting matter. Investigating for charged charmonium-like states is one of the most promising ways of searching exotic mesons since they must contain at least four quarks and thus cannot be a conventional hadron. 

Recently, for the first time, the BESIII Collaboration reported the charged strange hidden charmonium-like structure $Z_{cs}(3985)$ near the $ D^{-}_s D^{*0}$ and $ D^{*-}_s D^{0}$ mass thresholds in the $ K^+ $ recoil-mass spectrum for events collected at $ \sqrt{s} = 4.681 $ GeV in the processes of $e^+ e^- \rightarrow K^+ (D^{-}_s D^{*0}+ D^{*-}_s D^{0}) $ \cite{Ablikim:2020hsk}. The significance was estimated to be $ 5.3\sigma $. This discovery could maintain some unique hints to uncover the secrets of charged exotic $ Z $ structures. 
This new hadronic structure is assigned in the class of exotic state as the strange partner of $ Z_c(3900) $ and studied in many different models in the literature in the molecular and tetraquark scenarios \cite{1831062,1831033,1831047,Wang:2020kej,Meng:2020ihj,Liu:2020nge,Wan:2020oxt, Sun:2020hjw,Chen:2020yvq,Wang:2020rcx,Azizi:2020zyq,Wang:2020iqt,1832695}. Its mass and width are defined in experiment as:
\begin{eqnarray}
M_{Z_{cs}}&=&3982.5^{+1.8}_{-2.6}\pm 2.1~\mathrm{MeV},\notag\\
\Gamma_{Z_{cs}}&=&12.8^{+5.3}_{-4.4}\pm 3.0~\mathrm{MeV}.
\end{eqnarray}
Meanwhile, the features of matter under extreme conditions of high temperatures and/or densities have attracted the curiosity of high energy physicists ~\cite{Ayala:2020rmb,Ayala:2016vnt,Fu:2019hdw,Zhao:2020nwy,Irikura:2020}. QCD, the theory of strong interactions, expects that nuclear matter undergoes a phase transition from a state of deconfined quarks and gluons forming a new state of matter, named as the quark-gluon plasma (QGP), at a critical temperature $ T_c \cong 155$ MeV $ (\sim 10^{12}K) $ \cite{Aoki:2006br,Andronic:2017pug,Steinbrecher:2018phh,Fischer:2018sdj} which is in excellent agreement with the freeze-out temperature for hadrons measured by the ALICE collaboration at LHC producing  $ ^{4}\textrm{He} $ and $ ^{4}\overline{\textrm{He}} $ nuclei in Pb–Pb collisions at $ \sqrt{s_{NN}}=2.76 $ TeV in the rapidity range $ | y |< 1 $ \cite{Acharya:2017bso,Floris:2014pta}. Nevertheless its short life time ($ 10 fm/c \approx 3 \times 10^{-23} s  $) and thermalization time
($ 0.2 fm/c \approx 7 \times 10^{-25} s $), which makes measurements harder are a big challenge for experimentalist.

The phases of QCD are characterized by a variety of condensates in which numerous particles interact with each other through strong force. The materialization of condensates lessens the energy of a system, and also condensates break symmetries in QCD. Chiral symmetry breaking (CSB) is identified by a non-vanishing chiral condensate $ \langle \bar{q}q \rangle $, here $ q $ is the quark field. Luckily, at extreme temperatures, it is predicted that quark masses are decreased from their effective mass values in hot medium to their bare ones and CSB is almost restored. Namely, the condensates depending on temperature and baryon density play a key role in the structure of hadrons \cite{Hatsuda:1992bv,Foka:2016vta,Bratkovskaya:2017gxq}. 

Recreating longer-lived QGP as well as a large number and assortment of particles in laboratory conditions can provide us to explore the properties of QGP and also understand the QCD vacuum, confinement, and hadronization phase of the QGP. For a more precise interpretation of heavy-ion collision experiments, deviations of the hadronic parameters depending on temperature are vital and worth calculating.  From the theoretical point of perspective, these results have delivered some surprising and stimulating new theoretical studies of hot matter \cite{Lerambert-Potin:2021ohy,Grefa:2021qvt,Pinkanjanarod:2020mgi}. 

Charm and bottom quarks are excellent probes of the hot and dense state of deconfined quarks and gluons. Heavy quarks are created at the initial stages of the hard-scattering collisions and interact with the constituents of the newly produced QGP through both elastic and inelastic processes. These quarks, which can be studied through their decays into leptons, lose energy while propagating through the QGP medium. The formation of a QGP phase also lets them move freely and recombine to produce exotic states. They can coalesce to create standard and possibly exotic bound states at the end of the QGP phase. 

On the other hand, although a coherent picture
of collision dynamics is emerging, finding signatures of QGP remains unclear. Probably verification of QGP formation will
not come from a unique signal, and evidence based on
well-focused observations will have to
be collected. Some signatures supporting the creation of the QGP have been reported;
suppression (and regeneration) of heavy
quarkonia, jet-quenching, the non-viscous flow, radiation of
photons and dileptons. 

Depending on these ideas, the purpose of this article is to evaluate the mass, meson-current coupling constant  and decay width of $Z_{cs}(3985)$ assuming it has four-quark content $ [\bar{c}cu\bar{s}] $ including quark, gluon and quark-gluon mixed condensates up to
dimension six using the QCD Sum Rule (QCDSR) approach at finite temperature.
In this case, the vacuum condensate expressions are replaced with
their thermal condensates. This analysis can give us some hints on
the nature of the $Z_{cs}(3985)$ and also provide insights into the nature of the produced hot and dense matter which is predicted to exist in the initial stages of the universe and also in the core of neutron stars \cite{Baym:2017whm}.

The paper is organized as follows. The Thermal QCDSR (TQCDSR) approach is introduced employed in our calculations in Section \ref{sec:two}. Numerical analysis of mass and meson-current coupling constant of $Z_{cs}(3985)$ and its b-partner (after that we will use $Z_{cs}$ and  $Z_{bs}$ for shortness) is discussed in Section \ref{sec:Num}. In the next section, the decay width of $Z_{cs}(3985)$ is evaluated. After summarizing in Section \ref{result}, we present the explicit form of the two-point thermal spectral densities $\rho^{\mathrm{QCD}}(s,T)$ which are obtained from the TQCDSR theory in Appendix \ref{App}.

\section{THEORETICAL FRAMEWORK for two-point correlator}\label{sec:two}

The QCD sum rules technique is a successful and powerful non-perturbative method \cite{Shifman,Reinders:1984sr}, which is widely applied to study the mass spectra and decay properties of hadrons. To find the variations of mass and meson-current coupling of $Z_{cs}$ with increasing temperature, we adopt the QCDSR formalism to TQCDSR. We start the calculation by writing
down the correlation function \cite{Bochkarev:1985ex}:
\begin{equation}\label{eq:CorrF1}
\Pi_{\mu\nu}(q,T)=i\int d^{4}x~e^{iq\cdot x}\langle\Omega|
\mathcal{T}\{\eta_{\mu}(x) \eta_{\nu}^{\dag}(0)\}|\Omega\rangle,
\end{equation}
where $\mathcal{T}$ represents the time ordering operator, $\Omega $ symbolizes the thermal medium, $T$ is the temperature and $\eta_{\mu}(x)$ is the interpolating current
accompanying to resonance $Z_{cs}$.

To derive the TQCDSR we start to compute the
correlation function in connection with the physical degrees of
freedom. The correlation function is expressed by saturating via a complete set of states with the same quantum number $ J^{P}=1^{+} $ of $Z_{cs}$ state and then Eq.~(\ref{eq:CorrF1}) is integrated for $x$:
\begin{equation}
\Pi^{\mathrm{Phys}}_{\mu\nu}(q,T)=\frac{\langle
\Omega|\eta_{\mu}|Z_{cs}(q)\rangle\langle Z_{cs}(q)|\eta_{\nu}^{\dagger
}|\Omega\rangle}{m_{Z_{cs}}^{2}(T)-q^{2}}+ \ldots,
\end{equation}
where $m_{Z_{cs}}(T)$ is the temperature-dependent ground
state mass of axial-vector state $Z_{cs}$ and three dots indicate the
higher states and continuum. The definition of the matrix element of  temperature-dependent meson-current coupling constants is:
\begin{equation}\label{eq:Res1}
\langle\Omega|\eta_{\mu}|Z_{cs}(q)\rangle=\lambda_{Z_{cs}}(T)m_{Z_{cs}}(T)\varepsilon_{\mu},
\end{equation}
here $\varepsilon_{\mu}$ is the
polarization vector. So the correlation function for the physical side can be written
concerning the thermal ground state mass and meson-current coupling constant in the form below:
\begin{eqnarray}\label{eq:CorM}
\Pi_{\mu \nu }^{\mathrm{Phys}}(q,T)&=&\frac{m_{Z_{cs}}^{2}\left(T\right)\lambda_{Z_{cs}}^{2}\left(T\right)}{m_{Z_{cs}}^{2}\left(T\right)-q^{2}} \bigg (-g_{\mu\nu}+\frac{q_\mu q_\nu} {m_{Z_{cs}}^{2}\left(T\right)} \bigg )\notag\\
&+& \ldots.
\end{eqnarray}
In our computations, the chosen structure for both physical and QCD parts of the correlator is $ (g_{\mu \nu}) $ to obtain the TQCDSR for the mass and meson-current coupling constant. Then isolating ground state contributions from the higher resonances and continuum states
by taking derivative, namely using Borel transformation, the
physical side is determined as:
\begin{eqnarray}\label{eq:CorBor}
\mathcal{\widehat{B}}(q^{2})\Pi^{\mathrm{Phys}}(q^2,T)= m_{Z_{cs}}^{2}(T)\lambda_{Z_{cs}}^{2}(T)~e^{-m_{Z_{cs}}^{2}(T)/M^{2}},
\end{eqnarray}
here $M$ is the Borel parameter in the QCDSR model.

The next step is to determine the QCD part in which the
correlation function is expressed with the quark and gluon
degrees of freedom. First, we choose the
concerned current for the  $Z_{cs}$ state with $ J^{P}=1^{+} $ constructed in tetraquark picture as:  
\begin{eqnarray}\label{curraxial}
\eta_{\mu}(x)=i\epsilon_{abc}\epsilon_{dec}\big[\big(s_{a}^{T} (x) C\gamma_5
c_{b}(x)\big)\big(\overline{u}_{d}(x)\gamma_{\mu} C\overline{c}^{T}_e(x)
\big)\big],~
\end{eqnarray}
here $ \epsilon_{abc} $ and $\epsilon_{dec} $ are anti-symmetric Levi-Civita symbols, $a,b,c,d,e$ are color indices,  and $C$ is the charge conjugation matrix.

The QCD part of the correlation function
$\Pi_{\mu \nu }^{\mathrm{QCD}}(q,T)$ can be
described, as usual with a dispersion integral:
\begin{equation}
\Pi^{\mathrm{QCD}}(q^2,T)=\int_{\mathcal{M}^2}^{\infty}\frac{\rho
^{\mathrm{QCD}}(s,T)}{s-q^{2}}ds,
\end{equation}
where $\mathcal{M}^2=(m_{u}+m_{s}+2m_{c})^2$, and the spectral density function $\rho^{\mathrm{QCD}}(s,T)$ is given by the imaginary part of the correlation function:
\begin{equation}\label{eq:rhoQCD}
\rho^{\mathrm{QCD}}(s,T)=\frac{1}{\pi}Im[\Pi^{QCD}].
\end{equation}
Having completed lengthy calculations the QCD side of the correlation function in terms of the heavy and light quark propagators reads:
\begin{eqnarray}\label{eq:pi}
&&\Pi_{\mu\nu}^{\mathrm{QCD}}(q,T)=-\frac{i}{2}\int d^{4}xe^{iq\cdot x}\epsilon_{abc}\tilde{\epsilon}_{dec}\epsilon_{a'b'c'}^{\prime}\tilde{\epsilon}_{d'e'c'}^{\prime}\notag \\
&& \times \Big( \mathrm{Tr}[\widetilde{S}_{c}^{e^{\prime}e}(-x)\gamma_{\nu} S_{u}^{d^{\prime}d}(-x)\gamma_{\mu}]\mathrm{Tr}[\widetilde{S}_{s}^{aa^{\prime}}(x)\gamma_{5}S_{c}^{bb^{\prime}}(x)\gamma_{5}]\notag \\ 
&&-\mathrm{Tr}[\widetilde{S}_{c}^{e^{\prime}e}(-x)
\gamma_{5}S_{u}^{d^{\prime}d}(-x)\gamma_{\mu}]\mathrm{Tr}[\widetilde{S}_{s}^{aa^{\prime}}(x)\gamma_{5}S_{c}^{bb^{\prime}}(x)\gamma_{\nu}]\notag \\ 
&&-\mathrm{Tr}[\widetilde{S}_{c}^{e^{\prime}e}(-x)\gamma_{\nu}S_{u}^{d^{\prime}d}(-x)\gamma_{5}]\mathrm{Tr}[\widetilde{S}_{s}^{aa^{\prime}}(x) \gamma_{\mu}S_{c}^{bb^{\prime}}(x)\gamma_{5}]\notag \\ 
&&+\mathrm{Tr}[\widetilde{S}_{c}^{e^{\prime}e}(-x)\gamma_{5}S_{u}^{d^{\prime}d}(-x)\gamma_{5}]\mathrm{Tr}[\widetilde{S}_{s}^{aa^{\prime}}(x)\gamma_{\mu}S_{c}^{bb^{\prime}}(x)\gamma_{\nu}] \Big)\notag \\ 
\end{eqnarray}
and for compactness we used the following notation in Eq.~(\ref{eq:pi}):
\begin{equation*}
\widetilde{S}^{aa'}(x) = CS^{aa'T}(x)~C.
\end{equation*}

By the way, at finite temperatures, the additional operators
arise in the short distance expansion of the product of two quark
bilinear operators since the failure of Lorentz
invariance with the preferred reference frame and spilling of the
residual $\mathcal{O}(3)$ symmetry, and accordingly the thermal heavy and light quark propagators include new terms. So we modify the vacuum condensates by their thermal averages. 

In the calculations we use the following definition of the thermal light quark propagator $S_{q}^{ij}(x)$ \cite{Azizi:2016ddw,Azizi:2014maa,Mallik:1997pq}:
\begin{eqnarray}\label{lightquarkpropagator}
&&S_{q}^{ij}(x) =i\frac{\slashed
x}{2\pi^{2}x^{4}}\delta_{ij}-\frac{
m_{q}}{4\pi^{2}x^{2}}\delta_{ij} -\frac{\langle \bar{q}q\rangle_T }{12}\delta_{ij} \nonumber \\&&-\frac{x^{2}}{192}%
m_{0}^{2}\langle \bar{q}q\rangle_T \Big[1-i\frac{m_{q}}{6}\slashed x \Big]%
\delta _{ij}+\frac{i}{3}\Big[\slashed x
\Big(\frac{m_{q}}{16}\langle \bar{q}q\rangle_T \nonumber \\&&
-\frac{1}{12}\langle u^{\mu} \theta _{\mu \nu }^{f} u^{\nu}\rangle
\Big)+\frac{1}{3}(u\cdot x)\slashed u \langle
u^{\mu}\theta _{\mu \nu }^{f} u^{\nu}\rangle
\Big]\delta _{ij}   \nonumber \\ &&\nonumber \\
&&-\frac{ig_{s}G _{ij}^{\alpha\beta}}{32\pi ^{2}x^{2}}
\Big(\slashed x \sigma _{\mu \nu }+\sigma _{\mu \nu }\slashed
x\Big)-i\delta_{ij}\frac{x^2 \slashed x \langle \bar{q}q
\rangle^2_T}{7776} g_s^2,
\end{eqnarray}
where $m_{q}$ is the light quark mass, $\langle
\bar{q}q\rangle_T $ denotes the temperature-dependent light quark condensate, $u_{\mu }$ is the four-velocity of hot matter, and $\theta_{\mu \nu }^{f}$ is the fermionic part of the energy
momentum tensor. Also the gluon condensate
depending on the gluonic part of the energy-momentum
tensor $\theta _{\lambda \sigma }^{g}$ is \cite{Mallik:1997pq}:
\begin{eqnarray}\label{TrGG}
&&\langle Tr^{c}G_{\alpha \beta }G_{\mu \nu }\rangle
=\frac{1}{24}(g_{\alpha \mu }g_{\beta \nu }-g_{\alpha \nu
}g_{\beta \mu })\langle G_{\lambda \sigma
}^{a}G^{a\lambda \sigma }\rangle   \notag   \\
&&+\frac{1}{6}\Big[g_{\alpha \mu }g_{\beta \nu }-g_{\alpha \nu
}g_{\beta \mu }-2(u_{\alpha }u_{\mu }g_{\beta \nu }-u_{\alpha
}u_{\nu }g_{\beta \mu }\notag \\
&&-u_{\beta }u_{\mu }g_{\alpha \nu }+u_{\beta }u_{\nu }g_{\alpha \mu })\Big]%
\langle u^{\lambda }{\theta }_{\lambda \sigma }^{g}u^{\sigma
}\rangle.
\end{eqnarray}
The heavy quark propagator $S_{Q}^{ij}(x)$ $ (Q=c,b) $ is described as in Ref. \cite{Mallik:1997pq}:
\begin{eqnarray}\label{eq:HeavyProp}
&&S_{Q}^{ij}(x)=i\int \frac{d^{4}k}{(2\pi )^{4}}e^{-ik\cdot x}\Bigg[ \frac{%
\delta _{ij}\Big( {\!\not\!{k}}+m_{Q}\Big)
}{k^{2}-m_{Q}^{2}}\nonumber \\
&&-\frac{gG_{ij}^{\alpha \beta }}{4}\frac{\sigma _{\alpha \beta }\Big( {%
\!\not\!{k}}+m_{Q}\Big) +\Big(
{\!\not\!{k}}+m_{Q}\Big)\sigma_{\alpha
\beta }}{(k^{2}-m_{Q}^{2})^{2}}\nonumber \\
&&+\frac{g^{2}}{12}G_{\alpha \beta }^{A}G_{A}^{\alpha \beta
}\delta_{ij}m_{Q}\frac{k^{2}+m_{Q}{\!\not\!{k}}}{(k^{2}-m_{Q}^{2})^{4}}+\ldots\Bigg],
\end{eqnarray}
where for the external gluon field $G_{ij}^{\alpha \beta}$, the
below short-hand notation is employed:
\begin{equation*}
G_{ij}^{\alpha \beta }\equiv G_{A}^{\alpha \beta}\lambda_{ij}^{A}/2,
\end{equation*}
here $\lambda_{A}^{ij}$ are Gell-Mann matrices, $i,\,j$ are
color indices and $A=1,\,2\,\ldots 8$ are the number of gluon flavours. The first term in Eq.~(\ref{eq:HeavyProp}) denote
the perturbative contribution to the heavy quark propagator and
the others are non-perturbative terms.

Then taking into account tensor structure of $\Pi_{\mu\nu}^{\mathrm{QCD}}(q,T)$
we can write: 
\begin{eqnarray}
\Pi_{\mu\nu}^{\mathrm{QCD}}(q^2,T) & = & \Pi_{0}^{\mathrm{QCD}}(q^{2},T)\frac{q_{\mu}q_{\nu}}{q^{2}} \notag\\ &+&\Pi_{1}^{\mathrm{QCD}}(q^{2},T)(-g_{\mu\nu}+\frac{q_{\mu}q_{\nu}}{q^{2}}),
\end{eqnarray}
where $ \Pi_{0}^{\mathrm{QCD}}(q^{2},T) $ and $\Pi_{1}^{\mathrm{QCD}}(q^{2},T) $ are invariant functions. According to the idea of the QCDSR, we should choose the same structures $(g_{\mu\nu})$
for the mass and the meson-current coupling constant sum rules in both $\Pi_{\mu\nu}^{\mathrm{Phys}}(q^2,T)$
and $\Pi_{\mu\nu}^{\mathrm{QCD}}(q^2,T)$.

Now using these definitions, transferring the continuum contribution to the QCD part, applying Borel transformation to both parts of the sum rules and equating them, thermal meson-current coupling constant sum rule for the axial-vector meson $Z_{cs}$ up to the dimension-six condensates is written as follows:
\begin{eqnarray}\label{eq:lamdaSR}
&&m_{Z_{cs}}^{2}\left(T\right)\lambda_{Z_{cs}}^{2}(T)e^{-m_{Z_{cs}}^{2}(T)/M^{2}}~~~~~~~~~~~~~~~~~~~~~~~~~~~~~~\notag\\
&&~~~~~~~~~~~~~~~~~~~~=\int_{\mathcal{M}^2}^{s_{0}(T)}ds\rho^{\mathrm{QCD}}(s,T)e^{-s/M^{2}}
\end{eqnarray}
and then taking the derivative of Eq.~(\ref{eq:lamdaSR}) in terms of $(-1/M^2)$ we reach the thermal mass sum rule of $ Z_{cs} $:
\begin{equation}\label{eq:massSR}
m_{Z_{cs}}^{2}(T)=\frac{\int_{\mathcal{M}^2}^{s_{0}(T)}dss\rho^{\mathrm{QCD}}(s,T)e^{-s/M^{2}}}{\int_{\mathcal{M}^2}^{s_{0}(T)}ds\rho^{\mathrm{QCD}}(s,T)e^{-s/M^{2}}},
\end{equation}
where $s_0(T)$ symbolize the thermal
continuum threshold parameter which separates the ground state from higher states. The next step is to carry out the numerical analysis to determine the values of hadronic parameters of resonance $ Z_{cs} $ and also replace $ c $ quark with $b$ quark to obtain the $b$-partner $ Z_{bs} $ of $ Z_{cs} $  in tetraquark assumption.

\section{Analysis of the mass and meson-current coupling constant of the
 $ Z_{cs} $ and $ Z_{bs} $ }
\label{sec:Num}

To get the values of mass and meson-current coupling constant of the
hidden-charm system  $ Z_{cs} $ in the TQCDSR approach, we require some parameters e.g.
quark masses, quark, gluon, and mixed vacuum and thermal condensates. The vacuum values of these input parameters are listed in
Table~\ref{tab:input}.  
\begin{table}[htbp]
\caption{Input parameters.} \label{tab:input}
\begin{tabular}{|c|c|}
\hline
Parameters                               & Values \\
\hline\hline
$m_{u}$                                  & $2.16^{+0.49}_{-0.26}\mathrm{MeV} $ \cite{Zyla} \\
$m_{d}$                                  & $4.67 {}^{+0.48}_{-0.17}\mathrm{MeV} $ \cite{Zyla} \\
$m_{s}$                                  & $93^{+11}_{-5}\mathrm{MeV} $ \cite{Zyla} \\
$m_{c}$                                  & $1.23 \pm 0.09~\mathrm{GeV} $ \cite{Eidemuller:2000rc,Zyla} \\
$m_{b}$                                  & $4.18^{+0.03}_{-0.02}\mathrm{MeV} $ \cite{Zyla} \\
$\langle 0| \bar{q}q |0\rangle $         & $(-275(5))^3~\mathrm{MeV}^3$ \cite{Gubler:2018ctz} \\
$\langle  0| \bar{s}s  |0\rangle $       & $(-296(11))^3~\mathrm{MeV}^3$ \cite{Gubler:2018ctz} \\
$\langle\frac{\alpha_sG^2}{\pi}\rangle $ & $0.028(3)~\mathrm{GeV}^4$\cite{Horsley:2012ra} \\
$m_0^2 $                                 & $(0.8\pm0.1) ~\mathrm{GeV}^2 $ \cite{Shifman,Reinders:1984sr}  \\
\hline
\end{tabular}
\end{table}

Further, we need temperature-dependent quark, gluon condensates, and also energy
density as well. Thermal quark condensates are obtained fitting
data from Ref.~\cite{Gubler:2018ctz}, which is consistent with the Lattice QCD data:
\begin{eqnarray}\label{eq:qbarqT}
\frac{\langle\bar{q}q\rangle_{T}}{\langle 0| \bar{q}q |0\rangle}=
\mu_1 e^{c_1 T}+\mu_2,
\end{eqnarray}
where $q$ denotes the $u$ or $d$ quarks while for the $ s $ quark
\begin{eqnarray}\label{eq:sbarsT}
\frac{\langle\bar{s}s\rangle_{T}}{\langle 0| \bar{s}s |0\rangle}=
\mu_3 e^{c_2 T}+\mu_4,
\end{eqnarray}
here $ c_1=\mathrm{0.040~MeV^{-1}}$, $c_2=\mathrm{0.516~MeV^{-1}}$,
$\mu_1$$=-6.534\times10^{-4}$, $\mu_2=1.015$,
$\mu_3=-2.169\times10^{-5}$, $\mu_4$$=1.002$ are
coefficients \cite{Azizi:2019cmj} and are trustworthy up to a temperature $T=180~\mathrm{MeV}$ and $%
\langle 0|\bar{q}q|0\rangle $ denotes the condensate of the
light quarks at $ T=0 $.

The gluonic and fermionic parts of the energy density can be
parametrized as in Ref.~\cite{Azizi:2016ddw} taking into account the Lattice
QCD data given in Ref.~\cite{Cheng:2007jq}:
\begin{eqnarray}\label{tetaf}
\langle u^\mu\mathrm{\theta}^{f}_{\mu\nu}u^\nu\rangle_T &=& (\tau_1
e^{c_3T}+\mu_5)~T^{4},
\end{eqnarray}
\begin{eqnarray}\label{tetag}
\langle u^\mu\mathrm{\theta}^{g}_{\mu\nu}u^\nu\rangle_T &=& (\tau_2
e^{c_4 T}-\mu_6)~T^{4},
\end{eqnarray}
where $\tau_1=0.009 $, $c_3=24.876~\mathrm{GeV^{-1}}$,
$\mu_5=0.024$, $\tau_2=0.091 $, $c_4=21.277~\mathrm{GeV^{-1}}$
and $\mu_6=0.731$ \cite{Azizi:2019cmj}.

Also the temperature-dependent gluon condensate $ \langle
G^2\rangle_{T} $ is defined as in Ref. \cite{Gubler:2018ctz}:
\begin{eqnarray}\label{delta}
\delta \Big \langle \frac{\alpha_{s}G^{2}}{\pi}\Big
\rangle_{T}&=&-\frac{8}{9}[ \delta T^{\mu}_{\mu}(T)-m_{u} \delta
\langle\bar{u}u\rangle_{T}\nonumber
\\ &-&m_{d} \delta \langle\bar{d}d\rangle_{T}-m_{s} \delta
\langle\bar{s}s\rangle_{T}],
\end{eqnarray}
where the vacuum subtracted values of the related quantities are
employed as 
\begin{eqnarray}
\delta f(T)\equiv f(T)-f(0) 
\end{eqnarray}
and 
\begin{eqnarray}
\delta T^{\mu}_{\mu}(T)=\varepsilon(T)-3p(T) ,
\end{eqnarray}
here $ p(T) $ is the pressure and $\varepsilon(T)$ is the
energy density. Considering recent Lattice evaluations
\cite{Bazavov:2014pvz,Borsanyi:2013bia} the fit function of
$ \delta T^{\mu}_{\mu}(T) $ as obtained in Ref. \cite{Azizi:2019cmj}:
\begin{eqnarray}\label{epsmines3p}
\frac{\delta T^{\mu}_{\mu}(T) }{T^{4}}&=&(\mu_7 e^{c_5 T}+\mu_8)
\end{eqnarray}
with $\mu_7=0.020$, $c_5=29.412~\mathrm{GeV^{-1}}$,
$\mu_8=0.115$. 
Moreover the following expression for the  temperature-dependent strong coupling  \cite{Kaczmarek:2004gv,Morita:2007hv} is taken into account in the calculations being  $\Lambda_{\overline{MS}}\simeq T_{c}/1.14$:
\begin{eqnarray}\label{geks2T}
g_s^{-2}(T)=\frac{11}{8\pi^2}\ln\Big(\frac{2\pi T}{\Lambda_{\overline{MS}}}\Big)+\frac{51}{88\pi^2}\ln\Big[2\ln\Big(\frac{2\pi
T}{\Lambda_{\overline{MS}}}\Big)\Big].
\end{eqnarray}
As for the temperature-dependent continuum threshold $s_0(T)$ belonging to $ Z_{cs} $ state is another
auxiliary parameter that can be defined as in the following form  \cite{Dominguez:2016roi,Borsanyi:2010bp,Bhattacharya:2014ara}:
\begin{eqnarray}\label{eq:sOT}
\frac{s_0(T)}{s_0(0)}= \bigg( \frac{\langle \bar{q}q\rangle_T}{\langle 0| \bar{q}q |0 \rangle}\bigg)^{2/3}.
\end{eqnarray}
Continuum threshold parameter $s_0(0)$ is not completely independent of the mass of the first excited state of $ Z_{cs} $. According to the
QCDSR formalism, the physical quantities shouldn't be connected
with the auxiliary parameters $M^2$ and $s_0$.  However $M^2$ and $s_0$  are susceptible to
the selection of the parameters of the theory. 

In the QCDSR method, OPE convergence points us the lower bound on $M^2$, and the pole contribution (PC) yields the upper bound, i.e the highest-dimensional condensates should contribute no more than $\sim20\%  $ to the QCD side while the continuum is less than $ 50\% $ of the total terms. 

In this context, the maximum allowed $M^{2}$ needs be fixed to obey the restriction
dictated on $\mathrm{PC}$.  At the maximum value of $M^{2}$ the constraint $\mathrm{PC}>0.2$ is typical for multiquark systems and we get:
\begin{equation}
\mathrm{PC}=\frac{\Pi (M^{2},\ s_{0})}{\Pi (M^{2},\ \infty )}=0.24,  \label{eq:PC}
\end{equation}
here $\Pi (M^{2},\ s_{0})$ is the Borel-transformed and subtracted
invariant amplitude $\Pi ^{\mathrm{OPE}}(p^{2})$.

To ensure the convergence of the $\mathrm{OPE}$, at the minimum limit of $M^{2}$ we use
the limitation $R\leq 0.15$. The lower bound of Borel window is defined from the convergence of the $\mathrm{OPE}$ by the ratio below:
\begin{equation}
R(M^{2})=\frac{\Pi ^{\mathrm{(Dim5+Dim6)}}(M^{2},\ s_{0})}{\Pi (M^{2},\ s_{0})}=0.15,
\label{eq:Convergence}
\end{equation}
here $\mathrm{ Dim5} $ and $ \mathrm{Dim6} $ show the contributions to the
correlation function of the sum of the last two terms in the
operator product expansion.

\begin{figure}[htbp]
\begin{center}
\includegraphics[width=8.7cm]{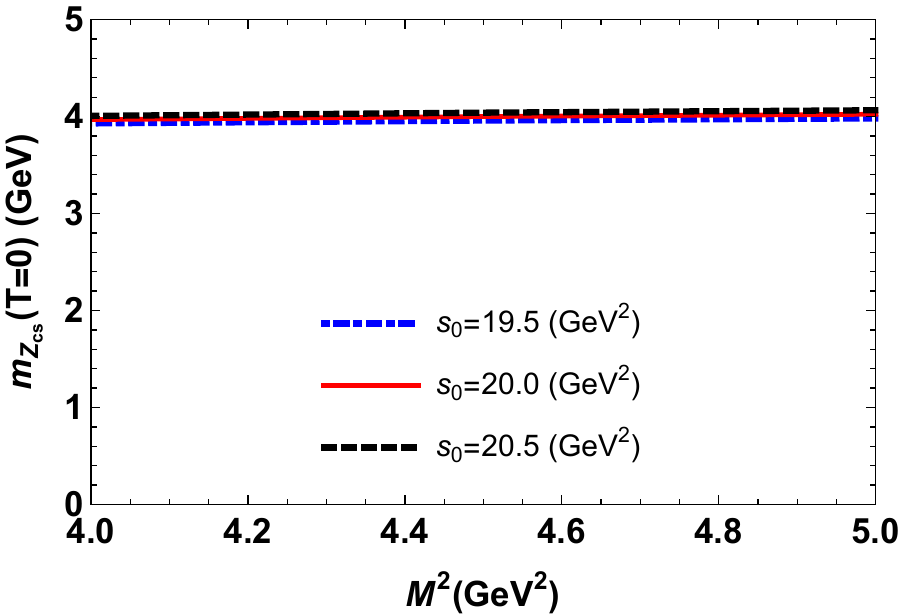}
\end{center}
\caption{The vacuum mass of the $Z_{cs}$ state versus Borel parameter $M^2$ for fixed continuum threshold values in tetraquark picture.} \label{mZcsMsq}
\end{figure}

Considering all these constraints, according to our analyses, the continuum threshold, and Borel parameters are fixed as follows for the $Z_{cs}$ resonance, respectively:
\begin{eqnarray}\label{eq:M2sOTvaluesZcs}
4~\mathrm{GeV^2} &\leq& M^2 \leq 5~\mathrm{GeV^2},\notag\\
19.5~\mathrm{GeV^2} &\leq& s_0~~ \leq 20.5~\mathrm{GeV^2},
\label{WR1}
\end{eqnarray}
and for the $Z_{bs}$ state as well:
\begin{eqnarray}\label{eq:M2sOTvaluesZbs}
13~\mathrm{GeV^2} &\leq&  M^2 \leq 15~\mathrm{GeV^2},\notag\\
122~\mathrm{GeV^2} &\leq& s_0 ~~ \leq 126~\mathrm{GeV^2}.
\label{WR2}
\end{eqnarray}
The philosophy of the QCDSR method dictates that the dependence of hadronic quantities on Borel parameter $M^2$ and continuum threshold $ s_0 $ should stay
steady in the selected working region. This means that we can obtain reliable results from the extracted sum rules. We see the stability of sum rules according to the model parameters drawing graphs. Here we only present a plot for the $Z_{cs}$ state in Figure \ref{mZcsMsq}.

In the end, the below results for $  Z_{cs} $ resonance in $ T=0 $ limit of the TQCDSR model is obtained: \\
\begin{eqnarray*}  \label{eq:ZcsResults}
m_{Z_{cs}}&=&3.996^{+0.068}_{-0.067}~\mathrm{GeV}, \\
\lambda_{Z_{cs}}&=& 0.72^{+0.06}_{-0.05}\times 10^{-2}~\mathrm{GeV^4},
\end{eqnarray*}
whereas we set a range for the state $ Z_{bs} $ as 
\begin{eqnarray*} 
m_{Z_{bs}}&=&(10.379 \sim 10.557)~\mathrm{GeV},\\ \lambda_{Z_{bs}}&=&(2.73 \sim 3.36)\times 10^{-2}~\mathrm{GeV^4},\\
\end{eqnarray*}
which are consistent with the experimental and theoretical estimations in Ref.~\cite{1831062,1831033,1831047,Wang:2020kej,Meng:2020ihj,Liu:2020nge,1832695,Azizi:2020zyq,Wang:2020iqt} within the limits of uncertainties \cite{Ablikim:2020hsk}.

\begin{figure}[htbp]
\begin{center}
\includegraphics[width=9cm]{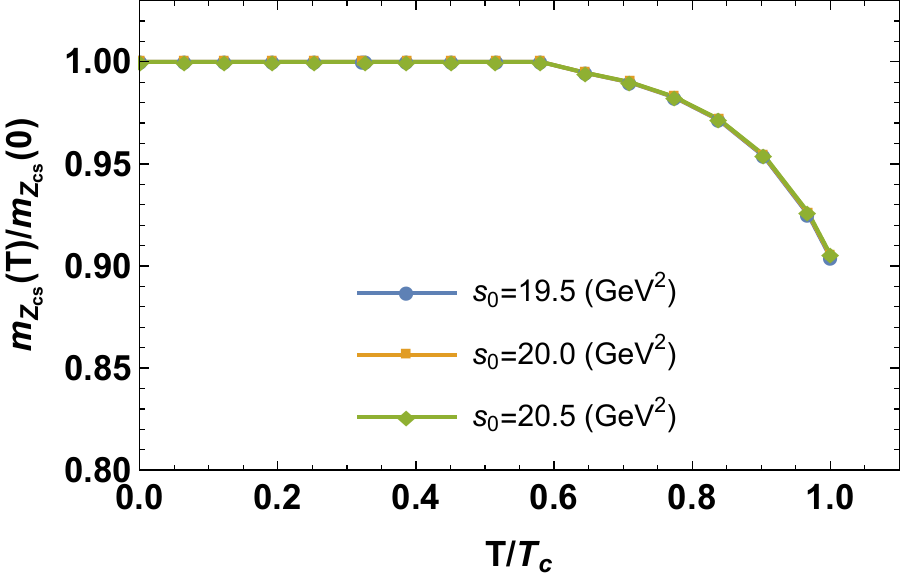}
\end{center}
\caption{The ratio of the temperature-dependent mass to vacuum  mass of the $Z_{cs}$ state, respectively in the tetraquark picture for fixed values of $s_0(0)$.} \label{fig2}
\end{figure}

\begin{figure}[htbp]
\begin{center}
\includegraphics[width=9cm]{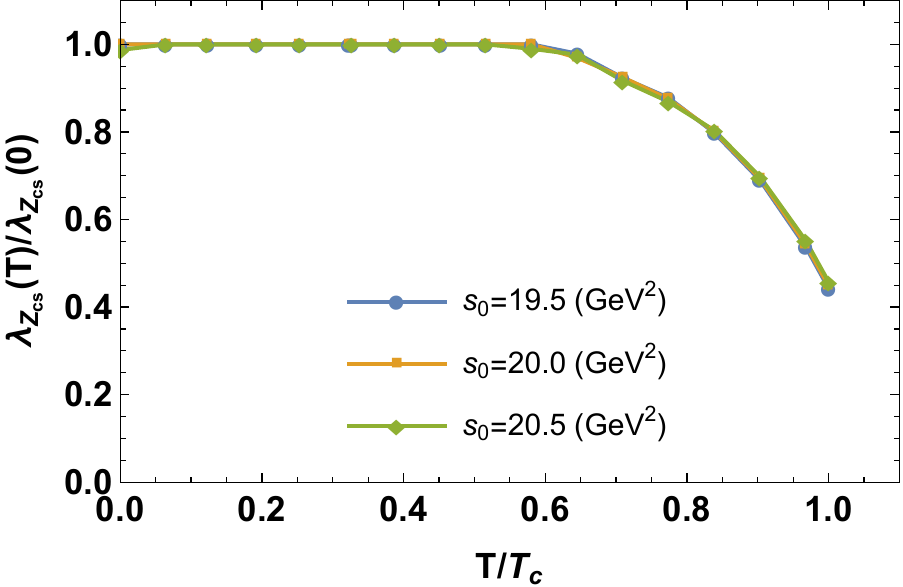}
\end{center}
\caption{The ratio of the temperature-dependent meson-current coupling constant to vacuum  meson-current coupling constant of the $Z_{cs}$ state, respectively in the tetraquark picture for fixed values of $s_0(0)$.} \label{fig3}
\end{figure}

Next, we define the modifications of mass and meson-current coupling constant of the $Z_{cs}$ in terms of temperature. In
this manner, the ratio of changing the mass and meson-current constant graphs are drawn as a
function of the temperature for the tetraquark assumption in Figure \ref{fig2} and \ref{fig3}, respectively. 
%
\section{Strong decays of the tetraquark $ Z_{cs}(3985)$}\label{width}	
%
The quark component of the newly observed resonance
$Z_{cs}$ should be $ [ c\bar{c}s\bar{u}] $ rather than the pure $ c\bar {c} $ since
it is a charged particle with strangeness and mass of this tetraquark is large enough to make
kinematically allowed the strong decay modes $ D_{s}^{-} D^{*0}/D_{s}^{*-} D^{0}$. In this part of the paper, we have discussed the details of the decays $Z_{cs}\rightarrow D_{s}^{-} D^{*0}/D_{s}^{*-} D^{0}$. 

$ \bullet $ We start with the first process $Z_{cs}\rightarrow D_{s}^{-} D^{*0}$. Initially, we need to compute the strong coupling corresponding to the vertex $Z_{cs}D_{s}^{-} D^{*0}$ which quantitatively defines strong
interactions between the tetraquark and two conventional
mesons. To obtain the QCD three-point sum rules for the related
coupling, we begin the calculation by writing the correlation function:
\begin{eqnarray}
\Pi _{\mu\nu}(p,p^{\prime },T) &=&i^{2}\int d^{4}xd^{4}ye^{i(p^{\prime}\cdot y-p\cdot x)}\langle\Omega|\mathcal{T}\{\eta_{\nu}^{D^{\ast 0}}(y)  \notag \\
&&\times \eta^{D_s^{-}}(0) \eta^{Z_{cs} }_{\mu}(x)\}|\Omega\rangle, \label{eq:CF2}
\end{eqnarray}
where $\eta^{Z_{cs} }_{\mu}(x)$, $\eta^{D_s^{-}}(0)$ and $\eta_{\nu}^{D^{\ast 0}}(y)$ symbolize the interpolating
currents for the tetraquark $Z_{cs}$ and mesons $D_s^{-}$ and $ D^{\ast 0}$, respectively. The four-momenta of the tetraquark $Z_{cs}$ and meson $D^{\ast 0 }$ are $p$ and $p^{\prime }$, respectively; and so the momentum of the meson $D_s^{-}$ is $q=p-p^{\prime}$. The current $\eta^{Z_{cs} }_{\mu}(x)$ is given by Eq.\ (\ref{curraxial}) in Section \ref{sec:two}, while the remaining two currents, we employ:
\begin{equation}
\eta_{\nu }^{D^{\ast 0 }}(y)=\overline{u}^{g}(y)i\gamma _{\nu }c^{g}(y),\ \
\eta^{D_s^{-}}(0)=i\overline{c}^{f}(0)\gamma _{5}s^{f}(0),  \label{eq:Curr3}
\end{equation}
where $g$ and $f $ are the color indices. Then, we apply the standard prescription of the TQCDSR technique and determine the correlation function $\Pi_{\mu\nu }(p,p^{\prime },T)$ using both physical
parameters of the hadrons involved in the process and quark-gluon degrees
of freedom. Isolating the ground-state contribution to the correlation
function in Eq.\ (\ref{eq:CF2}) from contributions of higher resonances and
continuum states 
for the physical side of the TQCDSR $\Pi _{\mu\nu }^{\mathrm{Phys}}(p,p^{\prime },T)$, we get:
\begin{eqnarray}
&&\Pi _{\mu\nu}^{\mathrm{Phys}}(p,p^{\prime },T)=\frac{\langle \Omega|\eta_{\nu}^{D^{\ast }}|D^{\ast}(p^{\prime })\rangle \langle \Omega|\eta^{D_s}|D_s(q)\rangle}{(m_{Z_{cs}}^{2}-p^2)(m_{D^{\ast}}^{2}-p'^2)}  \notag \\
&&\times \frac{\langle D_s(q)D^{\ast}(p^{\prime })|Z_{cs}(p)\rangle \langle Z_{cs}(p)|\eta_{\mu}^{\dagger Z_{cs}}|\Omega \rangle}{(m_{D_{s}}^{2}-q^{2})}+\ldots~~~~~  \label{eq:CF3}
\end{eqnarray}
To simplify this expression, we introduce the following matrix elements in terms of the meson's physical parameters:
\begin{eqnarray}
\langle \Omega|\eta^{D^{\ast}}_{\nu}|D^{\ast}(p')\rangle &=&f_{D^{\ast}}(T) m_{D^{\ast}}(T) \varepsilon'_{\nu},\   \notag \\
\langle Z_{cs}(p)|\eta_{\mu }^{Z_{cs}}|\Omega\rangle &=&\lambda_{Z_{cs}}(T) m_{Z_{cs}}(T)\varepsilon^{\ast}_{\mu },
\notag \\
\langle \Omega|\eta^{D_s}|D_s(q)\rangle &=&f_{D_s}(T)\frac{m^{2}_{D_s}(T)}{m_c+m_s}, \label{eq:Mel2}
\end{eqnarray}
here $m_{D^{\ast}}(T)$, $ m_{Z_{cs}}(T)$, $m_{D_s}(T)$ and $f_{D^{\ast}}(T)$, $\lambda_{Z_{cs}}(T)$, $f_{D_s}(T)$ are the temperature-dependent masses and decay constants of the mesons $D^{\ast} (2007)^0$, $Z_{cs}(3985) $ and $D_s(1968)^{-}$,
respectively. $ \varepsilon'_{\nu} $ and $\varepsilon _{\mu }^*$ are the
polarization vectors of the $D^{\ast }(2007)^{0}$ and $Z_{cs}(3985) $ states,
respectively. Then, we model $\langle D_s(q)D^{\ast}(p^{\prime },s^{\prime })|Z_{cs}(p,s)\rangle $ in Eq.\ (\ref{eq:Mel2}) as follows:
\begin{eqnarray}
\langle D_s(q)D^{\ast}(p^{\prime })|Z_{cs}(p)\rangle =g_{1}(T)[(p \cdot p^{\prime }) \notag\\
\times(\varepsilon ^{\prime *}\cdot\varepsilon )- (p \cdot \varepsilon ^{\prime *})(p^{\prime }\cdot\varepsilon )]\label{eq:Ver1}
\end{eqnarray}
denoting the strong coupling of the vertex $Z_{cs}(p)D_s(q)D^{\ast}(p^{\prime })$ with $g_{1}(T)$. Then, it is easy to get the physical part of the correlation function in Eq.\ (\ref{eq:CF3}):
\begin{eqnarray}
&&\Pi _{\mu \nu }^{\mathrm{Phys}}(p,p^{\prime },T) =\frac{
g_{1}(T)f_{D^{*}}(T) m_{D^*}(T)\lambda_{Z_{cs}}(T)m_{Z_{cs}}(T)}{(m_c+m_s)}~\notag \\
&&\times \frac{f_{D_s}(T)m_{D_s}^{2}(T)}{(m_{Z_{cs}}^{2}(T)-p^2)(m^2_{D^{*}}(T)-p^{\prime 2})(m_{D_s}^{2}(T)-q^{2})}\notag \\
&&\times\bigg[\frac{m_{Z_{cs}}^{2}(T)+m_{D^{\ast}}^{2}(T)-m_{D_s}^{2}(T)}{2}g_{\mu\nu}-p^{\prime }_{\mu }p_{\nu }\bigg] +\ldots 
\label{eq:Phys2}
\end{eqnarray}
The correlation function $\Pi _{\mu \nu }^{\mathrm{Phys}}(p,p^{\prime },T)$ ) contains the  two different Lorentz structures proportional to $p_{\mu}^{\prime} p_{\nu}$, $g_{\mu\nu }$ and one of which should
 be chosen to get the sum rules. We select the structure $
 g_{\mu\nu }$ to work with the invariant amplitude $\Pi ^{\mathrm{Phys}}(p^{2},p^{\prime 2},T)$.  Afterwards, we carry out the double Borel transformation of this amplitude over
variables $p^{2}$ and $p^{\prime 2}$. This operation allows us to arrive the physical side of the
sum rule. 

In order to get the other side, i.e. QCD side, of the three-point sum rule, we derive $\Pi
_{\mu\nu }(p,p^{\prime },T)$ in terms of the quark propagators:
\begin{eqnarray}
\Pi _{\mu\nu }^{\mathrm{QCD}}(p,p^{\prime },T)=i^{4}\int
d^{4}xd^{4}ye^{i(p^{\prime }\cdot y-p\cdot x)}\epsilon_{abc} \epsilon_{dec}
\notag \\ \times\mathrm{Tr}[ S_{c}^{ef}(x)\gamma _{5}
S_{s}^{fa}(-x)\gamma _{5}\widetilde{S}
_{c}^{gb}(y-x)\gamma _{\nu}\widetilde{S}_{u}^{dg}(x-y)\gamma_{\mu}].~\label{eq:CF4}
\end{eqnarray}\\
The correlation function $\Pi _{\mu\nu }^{\mathrm{QCD}}(p,p^{\prime },T)$ is
computed with dimension-6 accuracy, and has the same Lorentz structures as
$\Pi _{\mu\nu }^{\mathrm{Phys}}(p,p^{\prime },T)$. The double Borel
transformation $\mathcal{B}\Pi ^{\mathrm{QCD}}(p^{2},p^{\prime 2},T)$
unveils the second side of the sum rule. The Borel transformed and subtracted amplitude $\Pi ^{\mathrm{QCD}}(p^{2},p^{\prime 2},T)$ can be written depending on the spectral
density $\widetilde{\rho }(s,s^{\prime },T)$ which is proportional to
the imaginary part of $\Pi ^{\mathrm{QCD}}(p,p^{\prime},T)$,
\begin{eqnarray}
&&\Pi (\mathbf{M}^{2},\mathbf{\ s}_{0},T)=\int_{\mathcal{M}^{2}}^{s_{0}}ds%
\int_{m_{c}^{2}}^{s_{0}^{\prime }}ds^{\prime }\widetilde{\rho }(s,s^{\prime
},T)  \notag \\
&&\times e^{-s/M_{1}^{2}}e^{-s^{\prime }/M_{2}^{2}},  \label{eq:SCoupl}
\end{eqnarray}
where $\mathbf{M}^{2}=(M_{1}^{2},\ M_{2}^{2})$ and $\mathbf{s}_{0}=(s_{0},\
s_{0}^{\prime })$ represent the Borel mass and continuum threshold parameters,
respectively. The pair of parameters $ (M_{1},s_{0}) $ corresponds to the
initial tetraquark's channels, whereas $ (M_{2},s_{0}^{\prime }) $ depicts
the final-state meson. $ \mathcal{M}=(2m_c+m_s+m_u) $ and $\widetilde {\rho }(s,s^{\prime},T) $ are spectral densities computed as the imaginary parts of the corresponding terms in $ \Pi^{\mathrm{QCD}}_{\mu\nu }(p,p^{\prime},T) $. Next, by equating $\mathcal{B}\Pi ^{
	\mathrm{QCD}}(p^{2},p^{\prime 2},T)$ and Borel transformation of $\Pi ^{
	\mathrm{Phys}}(p^{2},p^{\prime 2},T)$, and doing continuum
subtraction, the sum rule for the coupling $g_{1}(T)$ is determined as:
\begin{eqnarray}
g_{1}(T)=\frac{2(m_{D_s}^{2}(T)-q^{2})}{f_{D^{\ast}}(T)m_{D^{\ast}}(T)\lambda_{Z_{cs}}(T) m_{Z_{cs}}(T)f_{D_s}(T)m_{D_s}^2(T)}
\notag \\
\times\frac{(m_{c}+m_s){\Pi }(\mathbf{M}^{2},\mathbf{\ s}_{0},q^{2})}{\big(m_{Z_{cs}}^{2}(T)+m_{D^{\ast}}^{2}(T)-m_{D_s}^{2}(T)\big)e^{-m_{Z_{cs}}^2/M_1^2}e^{-m_{D^*}^2/M_2^2}}\notag \\\label{eq:SRCoup}
\end{eqnarray}
Note that, the $g_{1}(T)$ is a function of $ T $, and also
rely on the Borel and continuum threshold parameters, but are not explicitly specified in Eq.\ (\ref{eq:SRCoup}) as arguments of $g_{1}$. After that, we introduce a new variable $Q^{2}=-q^{2}$ and denote the obtained function as $g_{1}(Q^{2})$.

The sum rule in Eq.\ (\ref{eq:SRCoup} ) includes mass and decay constant's temperature-dependence of the final mesons, so we need numerical values of these parameters. Therefore they are computed with the standard sum rule method and findings in vacuum are given in Table \ref{tab:Param2}. 
\begin{table}[h!]
\begin{tabular}{|c|c|}
		\hline\hline
		Parameters & Numeric Values~($\mathrm{GeV}$)\\ \hline\hline
		$m_{D^{0}}$ & $ 1.861^{+0.063}_{-0.062}$ \\
		$m_{D_{s}^{-}}$ & $ 1.966^{+0.060}_{-0.059}  $\\
		$m_{D^{*0}}$ & $2.005^{+0.051}_{-0.050} $ \\
		$m_{D_s^{*-}}$ & $2.071^{+0.025}_{-0.024} $  \\
		$f_{D^{0}}$ & $0.22^{+0.01}_{-0.01} $ \\
		$f_{D_{s}^{-}}$ & $  0.27^{+0.01}_{-0.01}$ \\	
		$f_{D^{*0}}$ & $0.23^{+0.01}_{-0.01}$ \\
		$f_{D_s^{*-}}$ & $0.27^{+0.01}_{-0.01}$\\ 
		\hline\hline
	\end{tabular}
	\caption{Obtained mass and coupling constant values of $D$ mesons at $ T=0 $  produced in the decays of tetraquark $Z_{cs}$.}
	\label{tab:Param2}
\end{table}
The following is the function that best fits the graphs we have drawn for the temperature dependencies of the mass and couplings:
\begin{equation}
	F_{n}(T)=A_{n}\mathrm{e}^{\frac{T}{B_{n}}}	+C_{n} ,
	\label{eq:FitF}
\end{equation}
here $F_{n}$, $A_{n}$, $B_{n}$ and $C_{n}$ are fitting parameters. 
Numerical analysis lets us fix these parameters as in Table \ref{tab:coeff1}. 
\begin{table}[h!]
	\begin{tabular}{|c|c|c|c|}
		\hline
$ F_{n} $		& $ A_{n}(\mathrm{GeV})$   & $ B_{n}(\mathrm{GeV}) $ & $ C_{n}(\mathrm{GeV}) $ \\ 
		\hline\hline
		$m_{D_{s}^{-}}$	&$ -1.375 \times10^{-4} $  &$ 0.021$                 &$ 1.970 $  \\ 
		$m_{D^{*0}}$    &$ -1.135\times10^{-4}  $  &$ 0.020$                 &$ 2.008 $    \\
		$m_{D_{s}^{*-}}$&$ -1.407 \times10^{-4} $  &$ 0.020$                  &$ 2.077 $  \\ 
		$m_{Z_{cs}}$    &$-2.206\times10^{-4}  $   &$ 0.020 $                &$ 4.001$    \\
		$m_{D^{0}}$     &$ -1.198\times10^{-4}  $  &$ 0.021$                 &$ 1.865 $    \\   
		\hline
		& $ A_{n}(\mathrm{GeV})$ & $ B_{n} (\mathrm{GeV})$           & $ C_{n} (\mathrm{GeV} ) $ \\ 
		\hline\hline
		$f_{D_{s}^{-}}$                 &$ -4.370\times10^{-5} $   &$ 0.021 $  &$ 0.271 $  \\
		$f_{D^{*0}}$                    &$ -3.068\times10^{-5}$    &$ 0.020 $  &$ 0.190 $  \\
		$f_{D_{s}^{*-}}$                &$ -3.690 \times10^{-5} $  &$ 0.021$   &$ 0.271 $ \\  
		$f_{D^{0}}$                     &$ -4.170\times10^{-5}  $  &$ 0.021$   &$ 0.224 $    \\    
		\hline
		& $ A_{n}(\mathrm{GeV^4})$ & $ B_{n} (\mathrm{GeV})$           & $ C_{n} (\mathrm{GeV^4} ) $ \\ 
		\hline\hline
		$\lambda_{Z_{cs}}$                    &$ -5.769\times10^{-4} $   &$ 0.023 $  &$ 0.007 $ \\
		\hline
		\end{tabular}
	\caption{Fit parameters for the mass and meson-current coupling constant of $ D_{s}^{-}, D^{*0}, D_{s}^{*-}, D^{0}$ and  $Z_{cs}$ states.}
	\label{tab:coeff1}
\end{table}\\\\\\
Then to continue the evaluation of $g_{1}(Q^{2})$,  we need to determine $\mathbf{M}^{2}
$ and $\mathbf{s}_{0}$. The limitations imposed on these auxiliary
parameters have been mentioned before. The working region for Borel mass and continuum threshold is the same range used in the mass and meson-current coupling constant calculation.
The decay width of the considered process should be computed using the strong
coupling at the $D_{s}^{-}$ meson's mass shell $Q^{2}=-m_{D_{s}}^{2}$, which is not
accessible to the sum rule calculations. We avoid this problem by
adopting a fitting procedure. Using the fit function below;
\begin{equation}
g_{fit}(Q^2)=\frac{g_{0}}{1-a\frac{Q^2}{m^2_{Z_{cs}}}+b\big(\frac{Q^2}{m^2_{Z_{cs}}}\big)^2}  \label{eq:Couplfit}
\end{equation}
where $  g_0=0.22,~a=-1.37,~b=-0.84 $ are the fit coefficients for the coupling $ g_1 $ which gives at the mass shell $Q^{2}=-m_{D_s}^{2}$ in vacuum:
\begin{equation}
g_{1}(-m_{D_s}^{2})=0.36\pm 0.02~\mathrm{GeV^{-1}} .  \label{eq:Coupl1}
\end{equation}
The decay width of $Z_{cs}\rightarrow D_s^{-}D^{*0}$ is extracted by
the following expression
\begin{eqnarray}
&&\Gamma_1 \left[ Z_{cs}\rightarrow D_{s}^{-} D^{\ast }(2007)^{0}\right] =\frac{
g_{1}^{2} (T) m_{D^{\ast }}^{2}(T)  }{24\pi}\notag\\
&&~~~~~~~~\times \xi ( m_{Z_{cs}}(T),m_{D^{\ast }}(T),m_{D_{s}^{-}}(T))\notag\\
&&~~~~~~~~\times
\Big[3+\frac{2 \xi^2 ( m_{Z_{cs}}(T),m_{D^{\ast }}(T),m_{D_{s}^{-}}(T))}{m_{D^{\ast }}^{2}(T)}\Big],\notag\\
\label{eq:DW1a}
\end{eqnarray}
where
\begin{equation}
\xi (a,b,c) =\frac{1}{2a}\sqrt{a^{4}+b^{4}+c^{4}-2(a^2 b^2+a^2 c^2+b^2 c^2) }.
\end{equation}
Employing the vacuum value of strong coupling from Eq.\ (\ref{eq:Coupl1}) and the $ m_{D^*} $ from Table \ref{tab:Param2}, the decay width of  $Z_{cs}\rightarrow D_{s}^{-} D^{\ast }(2007)^{0}$
\begin{equation}
\Gamma_1 \left[Z_{cs}\rightarrow D_{s}^{-} D^{\ast }(2007)^{0}\right] =(4.35\pm 0.44)\
\mathrm{MeV}\text{.}  \label{eq:DW1Numeric}
\end{equation}
$ \bullet $ The second process $\widetilde{Z}_{cs}\rightarrow D_{s}^{*-} D^{0}$ can be handled
in the same way as the first process.  However here, we utilize the following current expressions for the $ D^{ 0 } $ and $ D_s^{*-}$ mesons;
\begin{equation}
\eta_{\nu }^{D^{ 0 }}(y)=i\overline{u}^{g}(y)\gamma_{5}c^{g}(y),
\eta^{D_s^{*-}}(0)=\overline{c}^{f}(0)\gamma _{\nu }s^{f}(0),  \label{eq:Currsecdec}
\end{equation}
and introduce the new matrix elements:
\begin{eqnarray}
\langle \Omega|\eta^{D^0}|D^0(p')\rangle &=&\frac{f_{D^0}(T) m_{D^0}^2(T)}{m_c},
\notag \\\langle Z_{cs}(p)|\eta_{\mu }^{Z_{cs}}|\Omega\rangle &=&\lambda_{Z_{cs}}(T) ~m_{Z_{cs}}(T)~\varepsilon^{\ast}_{\mu },
\notag \\ \langle \Omega|\eta^{D_s^*}_{\nu}|D_s^*(q,\varepsilon^{'})\rangle &=&f_{D_s^*}(T)~m_{D_s^*}(T)~\varepsilon^{'}_{\nu }. \label{eq:Mel3}
\end{eqnarray}
By applying the standard procedures mentioned above for the first process, $\widetilde{\Pi }^{\mathrm{Phys}}(p,p^{\prime },T)
$ and $\widetilde{\Pi }^{\mathrm{OPE}}(p,p^{\prime },T)$ yield the sum rule
\begin{eqnarray}
g_2(q^{2})=\frac{(m_{D_s^*}^{2}(T)-q^{2})}{f_{D^0}(T)m_{D^0}^2(T)\lambda_{Z_{cs}}(T) m_{Z_{cs}}(T)f_{D_{s}^{\ast}}(T)m_{D_{s}^{\ast}(T)}}
\notag \\
\times\frac{2m_{c}\widetilde{\Pi}(\mathbf{M}^{2},\mathbf{s}_{0},q^{2})}{\big(m_{Z_{cs}}^{2}(T)+m_{D_s^{\ast}}^{2}(T)-m_{D^0}^{2}(T)\big)e^{-m_{Z_{cs}}^2/M_1^2}e^{-m_{D^0}^2/M_2^2}}.\notag \\
\label{eq:SRCoupl2}
\end{eqnarray}
Selecting the auxiliary parameters according to the same criteria as stated above and to determine the coupling constant $ g_2 $ at $ T=0 $, employing the fit function in Eq.\ (\ref{eq:Couplfit} ),  with coefficients $ g_0=0.24,~a=-1.52$, and $b=-0.92 $ at the mass shell $ Q^2=-m^2_{D_s^{*}} $ reads:
\begin{equation}
g_2( -m_{D_s^{*}}^{2}) =0.47\pm 0.03 ~\mathrm{GeV^{-1}}.
\end{equation}
The decay width of the second process is defined by the formula;
\begin{eqnarray}
&&\Gamma_2 [\widetilde{Z}_{cs}\rightarrow D_s^{*-}D^{0}]=\frac{g_2^{2}(T) m_{D_s^{*-}}^{2}(T)}{
24\pi }\notag \\
&&~~~~~~\times \xi \big( m_{\widetilde{Z}_{cs}}(T),m_{D_{s}^{*-}}(T),m_{D^{0}}(T)\big) 
\notag \\
&&\times \bigg(3+\frac{2\big[\xi ( m_{\widetilde{Z}_{cs}}(T),m_{D_{s}^{*-}}(T),m_{D^{0}}(T)) \big]^{2}}{m_{D_s^{*-}}^{2}(T)}\bigg),
\label{eq:DW2}
\end{eqnarray}
and our prediction for this decay channel is:
\begin{equation}
\Gamma_2 [\widetilde{Z}_{cs}\rightarrow D_s^{*-}D^{0}]=(7.65\pm 0.67)\ \mathrm{%
MeV}\text{.}  \label{eq:DW2a}
\end{equation}
As a result, we get the partial widths of these decays in the present section and using Eqs. (\ref{eq:DW1Numeric}) and (\ref{eq:DW2a}), the full width and mean lifetime of $ Z_{cs} $ are foreseen as:
\begin{eqnarray}
\Gamma^{\mathrm{full}}&=& (12.0\pm 0.8)~\mathrm{MeV}, \notag\\
\tau &=& (5.5\pm 0.5 )\times 10^{-19} s.
\label{lifetime}
\end{eqnarray}
Our result for the $ \Gamma^\mathrm{{full}} $ is consistent with the
measured value in BESIII \cite{Ablikim:2020hsk}. 
 
\begin{figure}[h!]
\includegraphics[width=8cm]{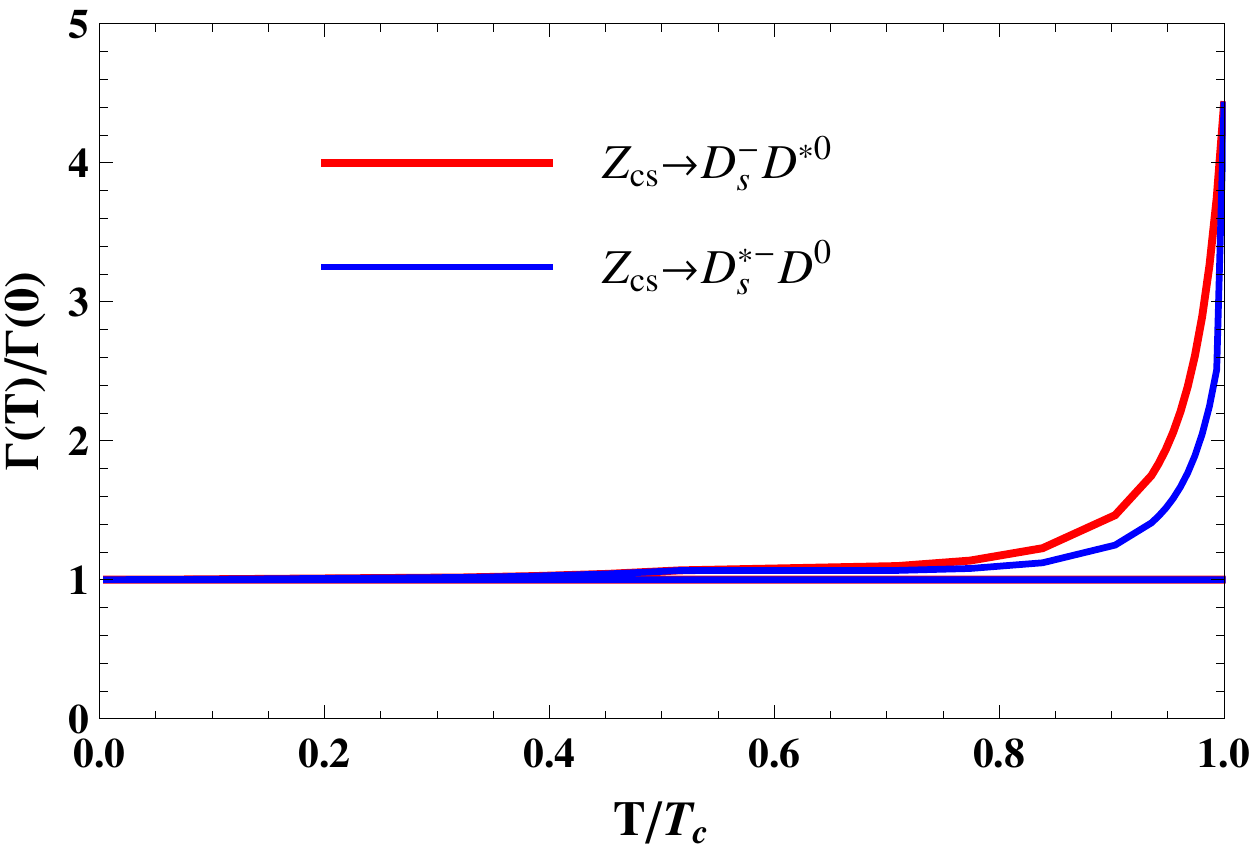}
\caption{Variation of the ratio of temperature-dependent decay width of $  Z_{cs}\rightarrow D_{s}^{-} D^{*0}/D_{s}^{*-} D^{0}  $ to its value in vacuum according to $ T/T_c $ for $ T_c=155 ~\mathrm{MeV}$.}
\label{fig:Gamma}
\end{figure}

The last step is to analyse the variation of partial decay widths in terms of temperature. We draw the $ \Gamma_{Z_{cs}}(T)/\Gamma_{Z_{cs}}(0) $ versus $ T/T_c $ in Figure (\ref{fig:Gamma}). As is seen from this figure decay width is dramatically increased with growing temperature. 
%
\section{\textbf{Summary and Results}}\label{result}

Collisions of heavy-ions in laboratory conditions allow us to create and investigate the strongly interacting matter in hot medium. Many facilities contribute to probing the chiral as well as the deconfinement phase transition from hadronic matter to the QGP and mapping out different domains of the QCD phase diagram. Among these experiments, Relativistic Heavy Ion Collider (RHIC) and Large Hadron Collider (LHC) energies form deconfined matter characterized by vanishing baryon densities and high temperatures consistent with lattice data. Mapping out the region of a first-order transition at large chemical potential is a primary aim of current and upcoming experimental programs.  Recently, RHIC and LHC have radically increased the energy levels that can be attained by heavy nuclei collisions at near-light speeds bringing them in line with those of the conditions in the early universe. In addition to these improvements, future experiments at the Facility for Antiproton and Ion Research (FAIR) and at the Nuclotron-based Ion Collider (NICA) will generate a wealth of data. 

 Also, the ALICE experiment is at the CERN LHC finalizing a major upgrade and will restart its operations with a new computing system in order to handle a data volume roughly $ 100 $ times larger than during the previous operational period in 2022. Further constraints can be set by future higher precision measurements during Run 3. The LHCC review of the ALICE $ 3 $ plans has begun and is expected to be concluded in March 2022 and deliver highly impactful physics results.  

However, in heavy-ion collisions, the QGP lifetime is $ (\sim 10 fm/c) $, and the hadrons would be erased from the spectrum in an extremely short time as seen from Eq.~(\ref{lifetime}). Hence it is almost impossible to work with a literally external probe in these collisions, whereas a hadron would continue to exist
although with changing mass in QGP medium. Alternatively, such an imaging method can be performed
using particles produced
during the parton-parton scatterings. So, searching the signals of QGP provide knowledge to quantitatively
understand the charmonium/bottomonium suppression of conventional and also exotic states \cite{Chen:2021akx,Abreu:2016qci,Abreu:2017pos} in heavy-ion collisions at high temperatures. This issue may be one of the focus areas of research in the near future regarding QGP signals.

Meanwhile, the newly observed resonance $ Z_{cs}(4000) $ with strangeness by LHCb collaboration got people to think whether they are the same state as the $ Z_{cs}(3985) $. But, according to our lifetime calculation of $ Z_{cs}(3985) $ detected in BESIII, has a narrower state from the $ Z_{cs}(4000) $ in LHCb and they should be different
states. This can be tested in future experiments and distinguish the two-state interpretation from the one-state scheme.

In this work, we propose a novel picture, i.e Thermal QCDSR, to understand the nature of $ Z_{cs}(3985) $ with $J^{P} = 1^+$ and also we reduce our results $ T=0 $ to compare with experimental and theoretical data in the literature. We also estimate the hadronic parameters of \textit{b}-partner of $ Z_{cs}(3985) $ which we hope to be detected in the near future experiments. According to our numerical evaluations, changes in mass and meson-current coupling constant values are fixed up to $T\cong100~\mathrm{MeV}$, but start to decrease after this point for both of them. At critical transition temperature the values of mass and meson-current coupling constant of  $ Z_{cs}(3985) $ change up to $10\%, 66\% $, respectively. The thermal width of the $ Z_{cs}(3985) $ meson (see Figure \ref{fig:Gamma}) exhibits an increase of roughly a factor 4.6 near $ T_c $. 

As a result, by looking at the numerical analysis we find that the resonance $ Z_{cs}(3985) $ can be well defined as a diquark-antidiquark candidate with quark content $ [\bar{c}cu\bar{s}] $. The variations on decay width of $ Z_{cs}(3985) $ will provide valuable input to our understanding of the heavy quark system in heavy-ion collisions. However, XYZ exotic states should be tested in more precise experimental data in the future and we need more experimental studies on the dominant decay channels of $ Z_{cs}(3985) $ to pin down its inner configuration. 

\appendix
\section{Thermal spectral density Functions}\label{App}
In this part, the results of our evaluations for the spectral
density is presented for the mass and meson-current coupling constant as a
function of the temperature belonging to the $ Z_{cs}(3985) $
resonance in the tetraquark picture using the following abbreviations as $\Theta$ is the step function (for brevity, we don't give the spectral
density belonging to the decay width here):
\begin{widetext}
\begin{eqnarray*}
L(s,x)&=&s x(1-x)-m_c^2, \\
L^{'}(s,x_1,x_2)&=&-\frac{(-1 + x_2) \Big[-s x_1 x_2 (-1 + x_1 + x_2) + m_c^2 (x_1 + x_2) (x_1^2 + x_1 (-1 + x_2) + (-1 + x_2) x_2)\Big]}{(x_1^2 + x_1 (-1 + x_2) + (-1 + x_2) x_2)^2},\\
\alpha&=&x_1^2 + x_1 (x_2-1) + x_2(x_2-1),\\
\beta&=&x_1 + x_2-1,\\
\zeta&=&x_1(x_1-1) +x_2 (x_1-1)  + x_2^2,\\
\eta&=&x_1 + x_2,\\
\end{eqnarray*}
and also we separate the thermal spectral density functions in terms of dimensions:
\begin{eqnarray} 
\rho^{\mathrm{QCD}}(s,T)&=&\rho^{\mathrm{pert.}}(s)+\rho^{\langle\bar{q}q\rangle}(s,T)+\rho^{G^2+\langle \theta_{00}\rangle}(s,T)
+\rho^{\langle qG q \rangle}(s,T)+\rho^{\langle\bar{q}q\rangle^2}(s,T).
\end{eqnarray}\\ 
The explicit form of spectral densities is performed with
the integrals over the Feynman parameters $x,x_1, x_2,$ and $ x_3$ as follows:
\begin{eqnarray}\label{eq:Rhopert}
	\rho^{pert}(s)&=&-\int_{0}^{1} dx_1 \int_{0}^{1-x_1} \frac{dx_2}{3072 \alpha^8 \beta \pi^6}\bigg\{\Big(-\alpha \eta m_c^2 + \beta s x_1 x_2 \Big)^2 \Big(12 \alpha^3 \eta m_c^3 m_s x_2 + 
	3 \alpha^2 \eta^2 m_c^4 x_1 x_2 - 48 \alpha^2 \beta m_c m_s s x_1 x_2^2 \notag\\ 
	&-& 26 \alpha \beta \eta m_c^2 s x_1^2 x_2^2 + 35 \beta^2 s^2 x_1^3 x_2^3 \Big)\bigg\}\Theta[L^{'}(s,x_1,x_2)],
\end{eqnarray}
\begin{eqnarray}\label{eq:dim3}
	\rho^{\langle\bar{q}q\rangle}(s,T)&=&\int_{0}^{1}  dx_1 \int_{0}^{1- dx_1} \frac{dx_2}{64 \alpha^6 \pi^4}\bigg\{15 \beta^3 m_s s^2 \langle \bar{s}s \rangle x_1^3 x_2^3 - 
	2 \alpha^3 \eta^2 m_c^5 \Big(2 \langle \bar{u}u \rangle x_1 + \langle \bar{s}s \rangle x_2 \Big) + 
	4 \alpha^2 \beta \eta m_c^3 s x_1 x_2 \Big(3 \langle \bar{u}u \rangle x_1 \notag\\
	&+& 2 \langle \bar{s}s \rangle x_2 \Big) - 
	2 \alpha \beta^2 m_c s^2 x_1^2 x_2^2 \Big(4 \langle \bar{u}u \rangle x_1 + 3 \langle \bar{s}s \rangle x_2 \Big) + 
	4 \alpha \beta m_c^2 m_s s x_1 x_2 \Big(-4 \beta \eta \langle \bar{s}s \rangle x_1 x_2 + \langle \bar{u}u \rangle \zeta^2 \Big) - 
	\alpha^2 \eta m_c^4 m_s \notag\\ 
	&\times& \Big(-3 \beta \eta \langle \bar{s}s \rangle x_1 x_2 + 4 \langle \bar{u}u \rangle \zeta^2 \Big)\bigg\}\Theta[L^{'}(s,x_1,x_2)],
\end{eqnarray}
\begin{eqnarray}\label{eq:dim4}	
	\rho^{G^2+\langle \theta_{00}\rangle}(s,T)&=& \int_{0}^{1} dx_1 \int_{0}^{1-x_1}dx_2 \Bigg\{ \frac{1}{2304 \alpha^6 \beta \pi^6}
	\Bigg[3 \beta x_2 \Big[4 \alpha^3 \eta m_c^3 m_s \Big(8 \beta \pi^2 \langle u^{\mu} \theta^f_{\mu \nu} u^{\nu} 
	\rangle + 
	g_s^2 \langle u^{\mu} \theta^g_{\mu \nu} u^{\nu} 
	\rangle x_1 \Big) + 
	\alpha^2 \eta^2 m_c^4 x_1\notag\\ 
	&\times&  \Big(64 \beta \pi^2 \langle u^{\mu} \theta^f_{\mu \nu} u^{\nu} 
	\rangle + 
	g_s^2 \langle u^{\mu} \theta^g_{\mu \nu} u^{\nu} 
	\rangle (3 x_1 - x_2)\Big) - 
	8 \alpha^2 \beta m_c m_s s x_1 \Big(16 \beta \pi^2 \langle u^{\mu} \theta^f_{\mu \nu} u^{\nu} 
	\rangle + 
	g_s^2  \langle u^{\mu} \theta^g_{\mu \nu} u^{\nu} 
	\rangle x_1\Big) x_2\notag\\
	&-& 4 \alpha \beta \eta m_c^2 s x_1^2 \Big(208 \beta \pi^2 \langle u^{\mu} \theta^f_{\mu \nu} u^{\nu} 
	\rangle  + 
	g_s^2 \langle u^{\mu} \theta^g_{\mu \nu} u^{\nu} 
	\rangle (4 x_1 - 5 x_2)\Big) x_2 + 
	5 \beta^2 s^2 x_1^3 \Big(192 \beta \pi^2 \langle u^{\mu} \theta^f_{\mu \nu} u^{\nu} 
	\rangle + 
	g_s^2 \notag\\ 
	&\times& \langle u^{\mu} \theta^g_{\mu \nu} u^{\nu} 
	\rangle (3 x_1 - 5 x_2)\Big) x_2^2 \Big] + 
	\langle \frac{\alpha_sGG}{\pi}\Big\rangle \pi^2 \Big[15 \beta^3 s^2 x_1^3 (5 x_1 - 9 x_2) x_2^3 -	12 \alpha \beta m_c m_s s x_1 x_2^2 \Big(4 (-1 + x_1)^2 x_1^2 \notag\\
	&+& 	4 (-1 + x_1) x_1 (-1 + 2 x_1) x_2 + 
	7 (-1 + x_1) x_1 x_2^2 + (-1 + 3 x_1) x_2^3 + x_2^4\Big) + 
	3 \alpha^2 m_c^3 \Big(m_s (-1 + x_1) x_1^3 (-12 \notag\\
	&+&  13 x_1) x_2 + 
	m_s (-1 + x_1) x_1^2 (-24 + 37 x_1) x_2^2 + 
	m_s x_1 \Big(12 + x_1 (-59 + 48 x_1)\Big) x_2^3 + m_s x_1 (-25 + 34 x_1) x_2^4\notag\\
	&+& m_s (-2 + 13 x_1) x_2^5 + 2 m_s x_2^6 \Big) + 
	3 \alpha^2 \eta^2 m_c^4 x_1 x_2 \Big(10 x_1^2 + (9 - 8 x_2) x_2 - 
	x_1 (9 + x_2)\Big) - 
	8 \alpha \beta \eta m_c^2 s x_1^2 x_2^2 \notag\\ 
	&\times& \Big(13 x_1^2 + (18 - 17 x_2) x_2 - x_1 (12 + 7 x_2)\Big)\Big]\Bigg] \Bigg\}
	\Theta[L^{'}(s,x_1,x_2)],		
\end{eqnarray}
\begin{eqnarray}\label{eq:dim5part1part2}
	\rho^{\langle qG q \rangle}(s,T)&=& \frac{m_c^2 m_o^2 m_s \langle \bar{u}u \rangle }{64 \pi^4}+\int_{0}^{1} dx_1 \int_{0}^{1-x_1}dx_2  \Bigg\{- \frac{\beta m_o^2 }{192 \alpha^5 \pi^4}\Big[-3 \alpha \beta \eta m_c^2 m_s \langle \bar{s}s \rangle  x_1 x_2 + 8 \beta^2 m_s s  \langle \bar{s}s \rangle x_1^2 x_2^2\notag\\
	&+& 3 \alpha^2 \eta m_c^3 \Big(2 \langle \bar{u}u \rangle x_1 + \langle \bar{s}s \rangle x_2 \Big) - 3 \alpha \beta m_c s x_1 x_2 \Big(3 \langle \bar{u}u \rangle x_1 + 2 \langle \bar{s}s \rangle  x_2 \Big)\Big]\Bigg\}\Theta[L^{'}(s,x_1,x_2)],
\end{eqnarray}
\begin{eqnarray}\label{eq:dim6part1part2}
	\rho^{\langle\bar{q}q\rangle^2}(s,T)&=&\frac{1}{1296 \pi^4}\Bigg[ \int_{0}^{1} dx \Bigg\{-108 m_c^2 \pi^2 \langle \bar{s}s \rangle \langle \bar{u}u \rangle + 
	m_c m_s \langle \bar{u}u \rangle \Big(g_s^2 \langle \bar{u}u \rangle (-1 + x) + 108 \pi^2 \langle \bar{s}s \rangle x \Big)  \Bigg\}\Theta[L(s,x)] \notag\\
	&+&\int_{0}^{1-x_1} dx_1 \int_{0}^{1-x_1-x_2}dx_2 \Bigg\{\beta^2 g_s^2 \big[  \langle \bar{s}s \rangle^2 + \langle \bar{u}u \rangle^2\big] x_1 x_2 (3 \alpha \eta m_c^2 - 8 \beta s x_1 x_2)\Bigg\}\Bigg]\Theta[L^{'}(s,x_1,x_2)].~~~~  
\end{eqnarray}\\\\
\end{widetext}

\end{document}